\documentclass[aps,twocolumn,prl,superscriptaddress]{revtex4-1}

\usepackage{amssymb}
\usepackage{amsmath}
\usepackage{graphicx}
\usepackage{epstopdf}
\usepackage{subfigure}
\usepackage{natbib}
\usepackage{epsfig}
\usepackage{amsfonts}
\usepackage{mathrsfs}
\usepackage{multirow}

\usepackage[colorlinks,linkcolor=blue,citecolor=blue,anchorcolor=blue]{hyperref}
\usepackage{dsfont,amsthm,amsbsy}
\usepackage{hyperref}
\hypersetup{
    colorlinks=true,
    linkcolor=blue,
    filecolor=magenta,
    urlcolor=cyan,
}

\begin{document}

\title{Less is more: Vacancy-engineered nodal-line semimetals}

\author{Fujun Liu}
\affiliation{Instituto de F\'{\i}sica, Universidade de Bras\'{\i}lia, Bras\'{\i}lia-DF, Brazil}

\author{Fanyao Qu}
\affiliation{Instituto de F\'{\i}sica, Universidade de Bras\'{\i}lia, Bras\'{\i}lia-DF, Brazil}

\author{Igor \v{Z}uti\'c}
\affiliation{Department of Physics, University at Buffalo, the State University of New York, Buffalo, NY 14260, USA}

\author{Mariana Malard}
\affiliation{Faculdade UnB Planaltina, Universidade de Bras\'{\i}lia, Bras\'{\i}lia-DF, Brazil}

\begin{abstract}
A nodal-line semimetal phase which is enforced by the symmetries of the material is interesting from fundamental and application standpoints. We demonstrate that such a phase of matter can be engineered by a simple method: Introducing vacancies in certain configurations in common symmorphic materials leads to nonsymmorphic polymorphs with symmetry-enforced nodal lines which are immune to symmetry-preserving perturbations, such as spin-orbit coupling. Furthermore, the spectrum acquires also accidental nodal lines with enhanced robustness to perturbations. These phenomena are explained on the basis of a symmetry analysis of a minimal effective two-dimensional model which captures the relevant symmetries of the proposed structures, and verified by first-principles calculations of vacancy-engineered borophene polymorphs, both with vanishing and with strong Rashba spin-orbit coupling. Our findings offer an alternative path to using complicated nonsymmorphic compounds to design robust nodal-line semimetals; one can instead remove atoms from a simple symmorphic monoatomic material.
\end{abstract}


\maketitle

\section{Introduction}

The study of degeneracies between energy bands in the spectrum of a system dates back to the early days of quantum mechanics~\cite{Neumann1929}. Since then, band degeneracies are featured in various physical phenomena, from signaling quantum phase transitions~\cite{Sachdev2001} to being the spectral signature of a semimetal~\cite{Murakami2007,Burkov2016,Armitage2018,Culcer2020}. According to the non-crossing rule by von Neumann and Wigner~\cite{Neumann1929}, energy bands generally avoid each other, but symmetries entail the possibility of band crossings. A band crossing is \emph{symmetry-enforced} if its existence is guaranteed by the underlying symmetry(ies) alone, and \emph{accidental} if the band crossing requires also tuning parameters of the material (e.g. a hopping energy, or any other microscopic parameter of the material). A special case occurs in three dimensions where the availability of sufficiently many tunable parameters (three momentum coordinates and one material parameter) leads to accidental band crossings even in the absence of symmetries~\cite{Murakami2007}. Symmetry-enforced band crossings arise from nonsymmorphic symmetries (point group transformations followed by a non-primitive translation), as first realized by Michel and Zak in 1999~\cite{MichelZak1999}, and later applied to the research on topological semimetals (TSMs)~\cite{Takahashi2017,Zhao2016,Schnyder2018,Bzdusek2016,Schoop2016,Alexandradinata2016,Young2015,Young2012}.

The nontrivial topology of band crossings in TSMs underpins a variety of phenomena, e.g. Fermi arcs~\cite{Wan2011} and chiral anomaly~\cite{Zyuzin2012}, and promise diverse technological applications, notably in topological quantum computing~\cite{Burkov2016}. TSMs with accidental band crossings are topologically protected only locally, being eventually spoiled by perturbations. In contrast, TSMs which have symmetry-enforced band crossings are endowed with global topological protection, i.e. they cannot be destroyed by symmetry-preserving perturbations~\cite{Zhao2016,Schnyder2018}.

Here we focus our attention on nodal-line semimetals (NLSMs) which are TSMs whose band crossings occur along lines in the Brillouin zone (BZ). Two-dimensional (2D) structures, e.g. hexagonal lattices~\cite{Xia2019} and honeycomb-kagome lattices~\cite{Lu2017}, have been predicted to realize NLSMs. However, the frailty to perturbations, particularly to spin-orbit coupling (SOC), of the resulting accidental nodal lines (NLs) hinders possible applications and use of proximity effects~\cite{Zutic2019}. Symmetry-enforced NLSMs are the natural candidates to overcome this challenge. Predicted realizations comprise 3D materials from hexagonal groups $P\bar{6}2c$, $P6_{1}22$, and $P6_{3}/m$~\cite{Zhang2018}.

We propose a simple mechanism for the realization of 2D symmetry-enforced NLSMs: Transform a monoatomic sheet with symmorphic symmetries into a nonsymmorphic one by removing atoms. Interestingly, besides symmetry-enforced NLs pinned at one edge of the BZ, our vacancy-engineered NLSMs exhibits also accidental NLs in the interior of the BZ which, unlike usual accidental NLs, survive under strong SOC. We have first proposed and outlined the basic principles of this scheme in a previous work~\cite{Liu2021} where we focused on the NLs in the interior of the BZ. Here we present a complete theoretical description of a 2D vacancy-engineered nonsymmorphic lattice which underpins the coexistence of symmetry-enforced and accidental NLs. While the latter are not directly wielded by the nonsymmorphic symmetry {\em per se}, they are a direct consequence of the proposed mechanism of attaining nonsymmorphicity out of vacancies. These accidental NLs can move and change shape inside the BZ, which might enable manipulation of momentum-dependent scattering processes and, hence, of various responses of the material. Moreover, our analysis of a 2D four-band effective model adds to the known proof of band-degeneracy enforcement for a 1D two-band model~\cite{Zhao2016,Schnyder2018}. The proposed mechanism of engineering nonsymmorphic 2D lattices from periodic configurations of vacancies thus opens an unexplored path to fundamental investigations and materials design of NLSMs.

\section{Material realizations}

\begin{figure}[htpb]
\includegraphics[width=90mm]{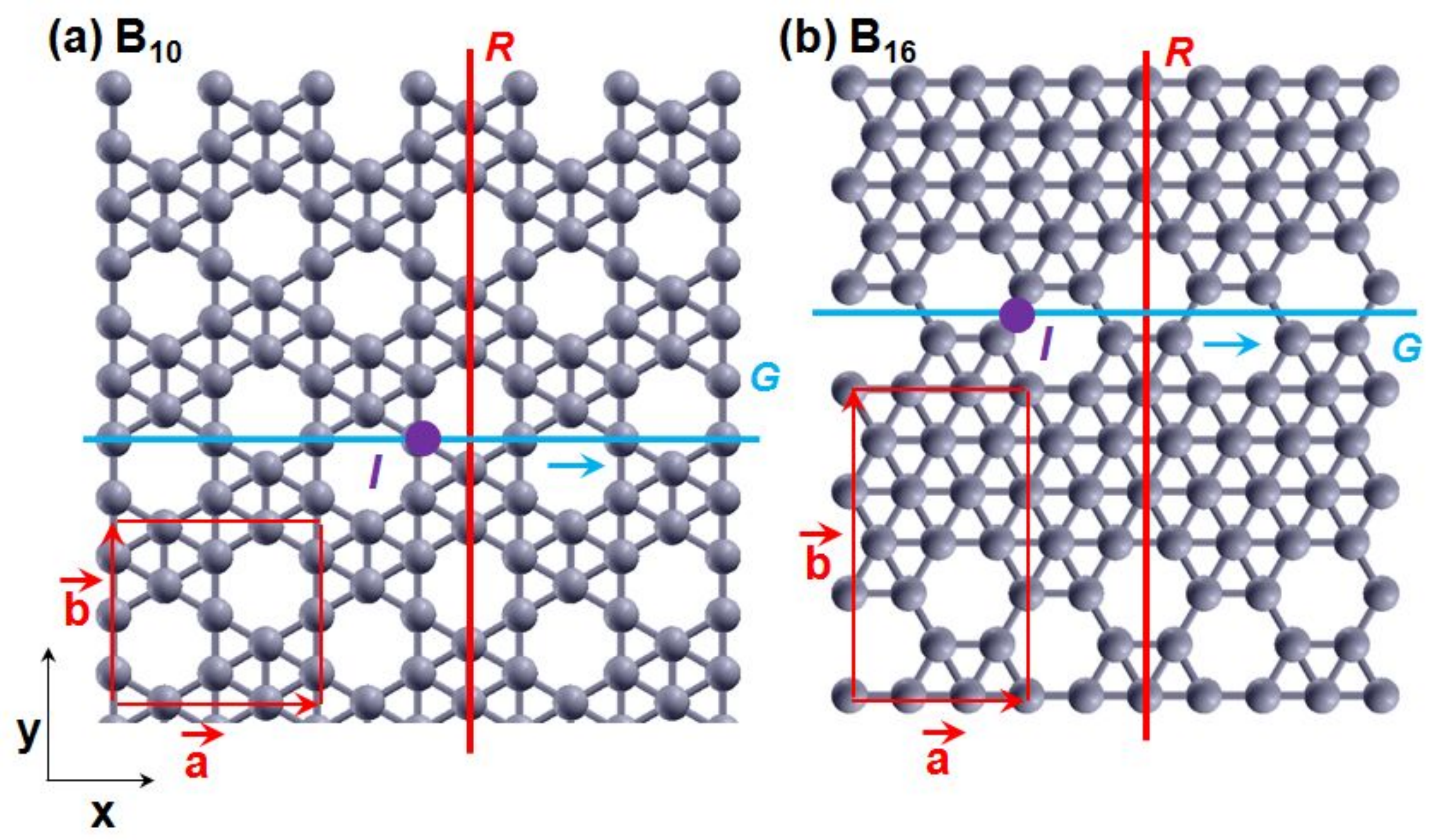}
\caption{(a) [(b)] Lattice structure of $\beta-$borophene B$_{10}$ [B$_{16}$] which has ten [sixteen] atoms in the unit cell defined by primitive vectors $\vec{a}$ and $\vec{b}$. B$_{10}$ [B$_{16}$] is obtained from pristine borophene by removing boron atoms from the center of hexagons which share one corner [side] along the $x$-direction. In both panels (a) and (b), the hollow hexagons form a string along the $x$-direction, with a zigzag profile in the $y$-direction. This basic feature yields a nonsymmorphic glide-plane symmetry $G$ (composed of a reflection plane running along the $x$-direction and a non-primitive translation by $\vec{a}/2$), a symmorphic reflection-plane symmetry $R$ perpendicular to $G$, and a symmorphic inversion-point symmetry $I$.}
\label{Fig1}
\end{figure}

As a concrete example, we illustrate our idea in borophene, a triangular 2D lattice of boron atoms which possesses polymorphs called $\beta$-borophenes~\cite{Tang2007,Yang2008,Ozdogan2010,Galeev2011,Penev2012,Mannix2015,Feng2016,Mannix2016,Zhou2021}. Here we propose and investigate the two stable $\beta$-borophenes shown on Figures~\ref{Fig1}(a)-(b), denoted as B$_{10}$ and B$_{16}$, where the subscripts refer to the number of atoms in the unit cell. B$_{10}$ and B$_{16}$ belong to the $pmg$ nonsymmorphic wallpaper group~\cite{Aroyo2016}. Pristine borophene without any vacancies is a symmorphic material; it possess two perpendicular reflection planes which entail the appearance of accidental Dirac cones in the spectrum, akin to those in graphene, silicene and germanene. Introducing vacancies at proper concentrations and configurations in pristine borophene yields that one of the reflection planes is replaced by a nonsymmorphic glide plane (c.f. Figs.~\ref{Fig1}(a)-(b)), and the Dirac cones give place to symmetry-enforced NLs. On top of that, the spectrum acquires also accidental NLs which are robust to strong Rashba SOC.

\section{Symmetry-enforced band degeneracies of a nonsymmorphic two-dimensional lattice}

\begin{figure}[htpb]
\centering
\includegraphics[width=88mm]{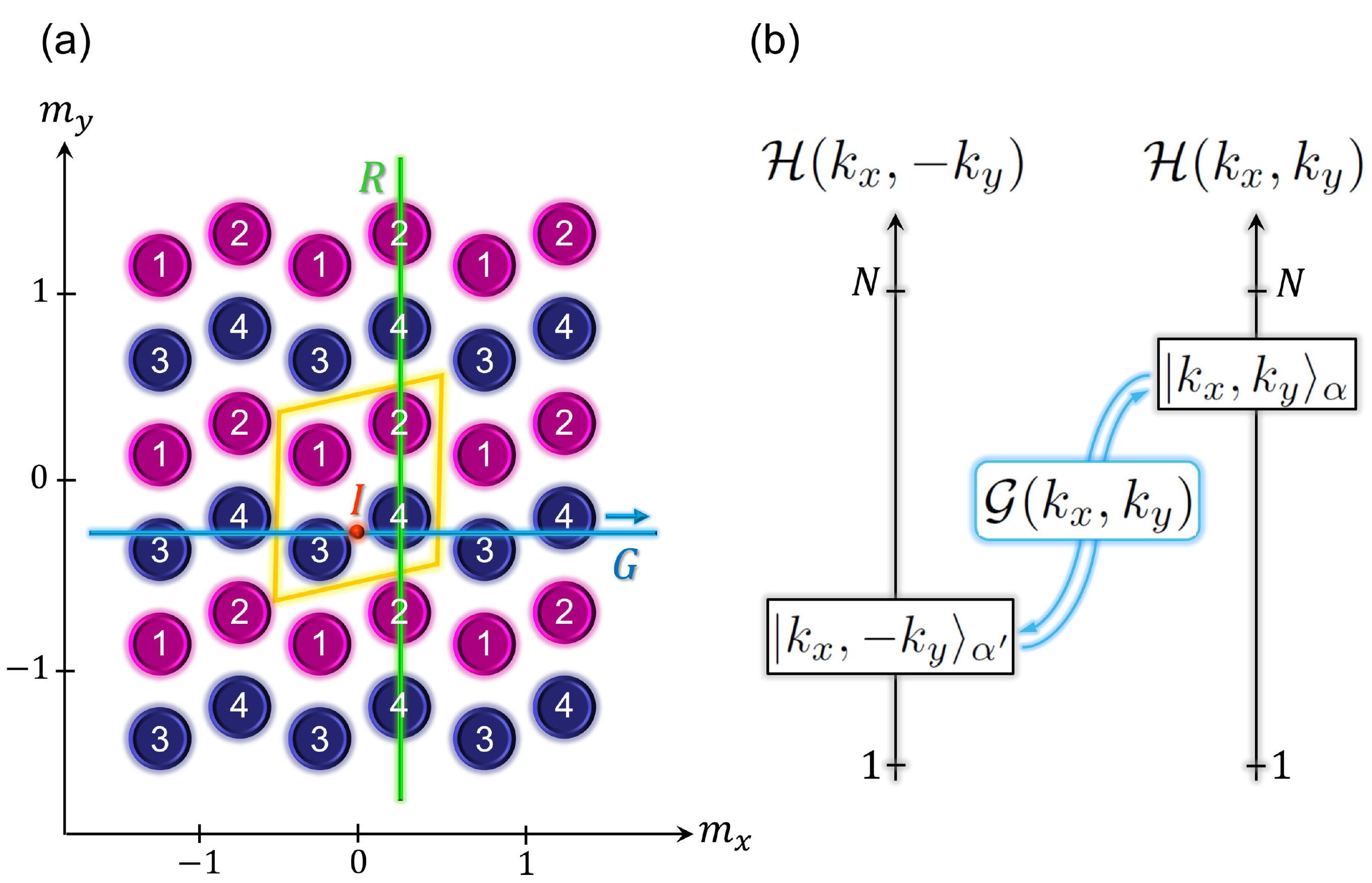}
\caption{(a) A minimal nonsymmorphic two-dimensional lattice, with the unit cell delineated by the yellow lines. The location of the unit cell is defined by $(m_{x},m_{y})$. The circles numbered from 1 to 4 represent two types of structures within the unit cell. In real $\beta-$borophenes B$_{10}$ and B$_{16}$ shown in Fig.~\ref{Fig1}(a)-(b), these structures are the hollow and filled pieces that make up the unit cell of those materials. The minimal lattice has the following spatial symmetries: a nonsymmorphic glide plane, $G$, composed of a reflection about the $m_{x}$-direction, followed by a nonprimitive translation by half of the length of the unit cell along the $m_{x}$-direction, a symmorphic reflection plane, $R$, about the $m_{y}$-direction, and a symmorphic inversion point, $I$, which takes a point $\vec{r}$ on the lattice to $-\vec{r}$. (b) The glide-plane symmetry ${\cal G}(k_{x},k_{y})$ transforms the $|k_{x},k_{y}\rangle_{\alpha}$ eigenstate of the Bloch Hamiltonian ${\cal H}(k_{x},k_{y})$ into the $|k_{x},-k_{y}\rangle_{\alpha'}$ eigenstate of ${\cal H}(k_{x},-k_{y})$, and vice-versa. $N$ is the number of bands ($N=4$ for the lattice depicted in (a)).}
\label{Fig2}
\end{figure}

In this section we carry out a symmetry analysis of a minimal lattice which captures the symmetries of B$_{10}$ and B$_{16}$ shown in Figs.~\ref{Fig1}(a)-(b). Figure~\ref{Fig2}(a) depicts a 2D lattice whose unit cell has four internal degrees of freedom, represented by the magenta and blue circles which are shifted along the $m_{y}$-direction. This shift mimics the profile of the filled and hollowed stripes in B$_{10}$ and B$_{16}$. The minimal lattice is invariant under a nonsymmorphic glide plane $G$, a symmorphic reflection plane $R$, and a symmorphic inversion point $I$. We now analyze how these symmetries constrain the band structure of the $4\times4$ Bloch Hamiltonian, ${\cal H}(k_{x},k_{y})$, of the lattice depicted in Fig.~\ref{Fig2}(a).

The invariance of the lattice with respect to the glide-plane transformation $G$ is manifest in the relation
\begin{equation} \label{GInvariance}
{\cal G}(k_{x},k_{y}){\cal H}(k_{x},-k_{y}){\cal G}^{-1}(k_{x},k_{y})={\cal H}(k_{x},k_{y}),
\end{equation}
where ${\cal G}(k_{x},k_{y})$ is the $4\times4$ matrix representation of $G$ in the basis constructed by the eigenstates of ${\cal H}(k_{x},k_{y})$. The $k_{x}$-dependance of ${\cal G}$ stems from the fractional translation along the $m_{x}$-direction, while the $k_{y}$-dependance originates from the shift of the glide plane from the center of the unit cell along the $m_{y}$-direction (c.f. Fig.~\ref{Fig2}(a)). So $G$ is an unusual symmetry which is both nonsymmorphic (along $m_{x}$) and off-centered~\cite{Yang2017,Malard2018} (along $m_{y}$). Such a glide plane differs from the one used to prove band-degeneracy enforcement in a 1D two-band model, the latter being a $2\times2$ matrix which depends on one momentum coordinate only~\cite{Zhao2016,Schnyder2018}.

\begin{figure}[htpb]
\centering
\includegraphics[width=88mm]{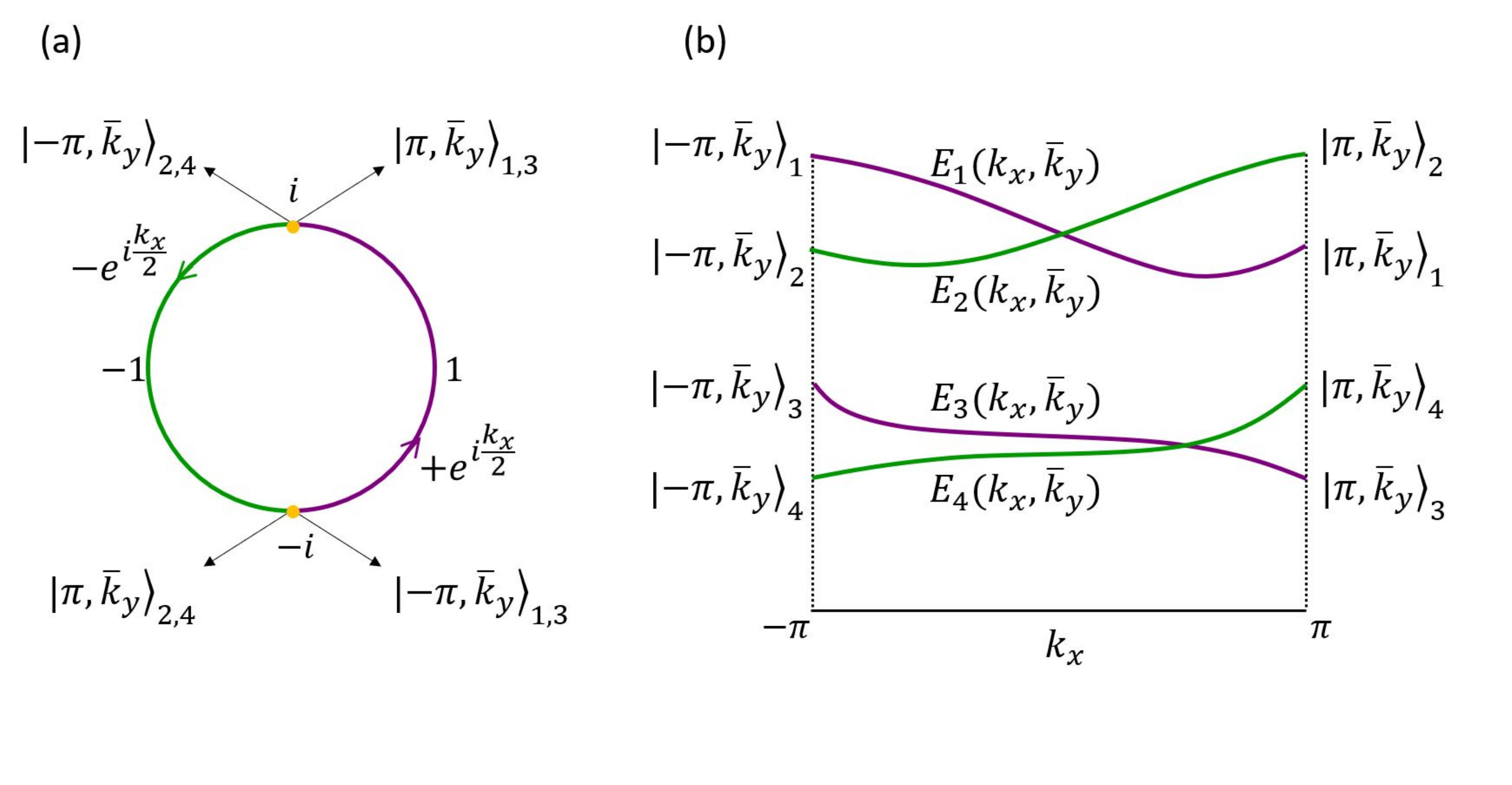}
\caption{(a) Doubly-degenerate eigenvalues $\pm e^{i\frac{k_{x}}{2}}$ of the glide-plane matrix winding around the half-unit circles on the complex plane, one pair of eigenvalues from $-i$ to $i$ through 1, and the other pair from $i$ to $-i$ through -1. The eigenstates $|\pm\pi,\bar{k}_{y}\rangle_{1,3}$ and $|\pm\pi,\bar{k}_{y}\rangle_{2,4}$ associated to the extreme points of the eigenvalue-trajectories are indicated. (b) The behavior of the eigenvalues of the glide-plane matrix implies that the eigenvalues of the Bloch Hamiltonian must cross pairwise at some value of $k_{x}$.}
\label{Fig3}
\end{figure}

We show (c.f. Appendix A) that ${\cal G}(k_{x},k_{y})|k_{x},-k_{y}\rangle_{\alpha'}\rightarrow|k_{x},k_{y}\rangle_{\alpha}$, where $|k_{x},k_{y}\rangle_{\alpha}$ ($\alpha=1,2,3,4$) is a Bloch eigenstate of ${\cal H}(k_{x},k_{y})$. This transformation between the negative-$k_{y}$ and positive-$k_{y}$ Bloch eigenspaces is illustrated in Fig.~\ref{Fig2}(b). On the lines $k_{y}=\bar{k}_{y}=0,\pm\pi$, ${\cal H}(k_{x},-\bar{k}_{y})={\cal H}(k_{x},\bar{k}_{y})$. Eq.~(\ref{GInvariance}) thus yields $[{\cal G}(k_{x},\bar{k}_{y}),{\cal H}(k_{x},\bar{k}_{y})]=0$ and, hence, $|k_{x},\bar{k}_{y}\rangle_{\alpha}$ are also eigenstates of ${\cal G}(k_{x},\bar{k}_{y})$. By constructing the matrix ${\cal G}(k_{x},k_{y})$ (c.f. Appendix A), we obtain the two-fold degenerate eigenvalues of ${\cal G}(k_{x},\bar{k}_{y})$: $\xi_{1,3}=\,\textrm{exp}(ik_{x}/2)$, and $\xi_{2,4}=-\,\textrm{exp}(ik_{x}/2)$. As $k_{x}$ swipes the BZ from $-\pi$ to $\pi$, the $(+)$-eigenvalues wind around the half-unit circle on the complex plane from $-i$ to $i$ through 1, while the $(-)$-eigenvalues wind from $i$ to $-i$ through -1, as illustrated on Figure~\ref{Fig3}(a). As a result, the eigenstates $|-\pi,\bar{k}_{y}\rangle_{1,3}$ and $|\pi,\bar{k}_{y}\rangle_{2,4}$ have the same ${\cal G}$-eigenvalue, $-i$, and the eigenstates $|\pi,\bar{k}_{y}\rangle_{1,3}$ and $|-\pi,\bar{k}_{y}\rangle_{2,4}$ have the same ${\cal G}$-eigenvalue, $i$. It follows that the associated pairs of ${\cal H}$-eigenvalues must cross at least once along the $k_{x}$-axis~\cite{Zhao2016} (when $k_{y}=\bar{k}_{y}$), as shown in Fig.~\ref{Fig3}(b).

To further clarify the origin of the band crossings depicted in Fig.~\ref{Fig3}(b), we analyze the simplest tight-binding model for the lattice shown in Fig.~\ref{Fig2}(a) in which only hopping between nearest-neighbor sites and on-site energies are considered. The entries of the corresponding Bloch Hamiltonian ${\cal H}(k_{x},k_{y})$ read
\begin{eqnarray}\label{epsilon}
\nonumber&\varepsilon&_{n,m}(k_{x},k_{y})\,=\,t_{n,m}+t_{m,n}^{\ast}+u_{n,m}e^{ik_{x}}+u_{m,n}^{\ast}e^{-ik_{x}}+\\
\nonumber+&v&_{n,m}e^{ik_{y}}+v_{m,n}^{\ast}e^{-ik_{y}}+w_{n,m}e^{i(k_{x}+k_{y})}+w_{m,n}^{\ast}e^{-i(k_{x}+k_{y})},\\
.
\end{eqnarray}
where $n,m=1,...,4$, and $t_{n,m}$, $u_{n,m}$, $v_{n,m}$, and $w_{n,m}$ denote the hopping energy from site $m$ to $n$ within the same unit cell, between neighboring unit cells along the $x$-direction, $y$-direction, and diagonal direction, respectively, and with $t_{n,n}=\mu_{n}$ the on-site energy. In Eq.~(\ref{epsilon}), only $u_{2,1}=t_{2,1}$, $u_{4,1}=t_{4,1}$, $u_{4,3}=t_{4,3}$ along $x$, $v_{1,3}=t_{1,3}$, $v_{2,3}=t_{2,3}$, $v_{2,4}=t_{2,4}$ along $y$, and $w_{2,3}=t_{2,3}$ along the diagonal are non-vanishing intercell hopping (c.f. Fig.~\ref{Fig2}(a)).

Imposing Eq.~(\ref{GInvariance}) (with ${\cal G}(k_{x},k_{y})$ given by Eq.~(\ref{G})), leads to Eqs.~(\ref{GPconstraints1})-(\ref{GPconstraints4}) (c.f. Appendix A) constraining the entries $\varepsilon_{n,m}(k_{x},k_{y})$ of ${\cal H}(k_{x},k_{y})$. A similar argument to the one used for the off-diagonal entry of a 1D two-band model~\cite{Zhao2016} can be applied here to show that Eq.~(\ref{GPconstraints2}) implies that $\varepsilon_{n,n+1}(k_{x},\bar{k}_{y})$ must vanish at some value of $k_{x}$. In a 1D two-band model, since there is only one off-diagonal entry, its vanishing is sufficient to guarantee a band crossing. In a multi-band model, on the other hand, the vanishing of only one of its off-diagonal entries is not a sufficient condition. For a 1D multi-band model, band crossings occur provided that the model has, on top of the nonsymmorphic symmetry, also chiral symmetry~\cite{Zhao2016}. Chiral symmetry means that the Bloch Hamiltonian admits an off-diagonal block form which, in turn, means that half of its entries are identical to zero. This is clearly not the case of our 2D four-band ${\cal H}(k_{x},k_{y})$ with entries given by Eq.~(\ref{epsilon}), which allows for hopping between any two intracell sites and on-site energies. This is particularly relevant for making a connection with a real 2D material in which hopping occurs in all directions. Therefore, the (nonsymmorphic + chiral)-symmetries argument designed for the 1D multi-band case is not applicable here.

\begin{figure*}[htpb]
\includegraphics[width=16cm]{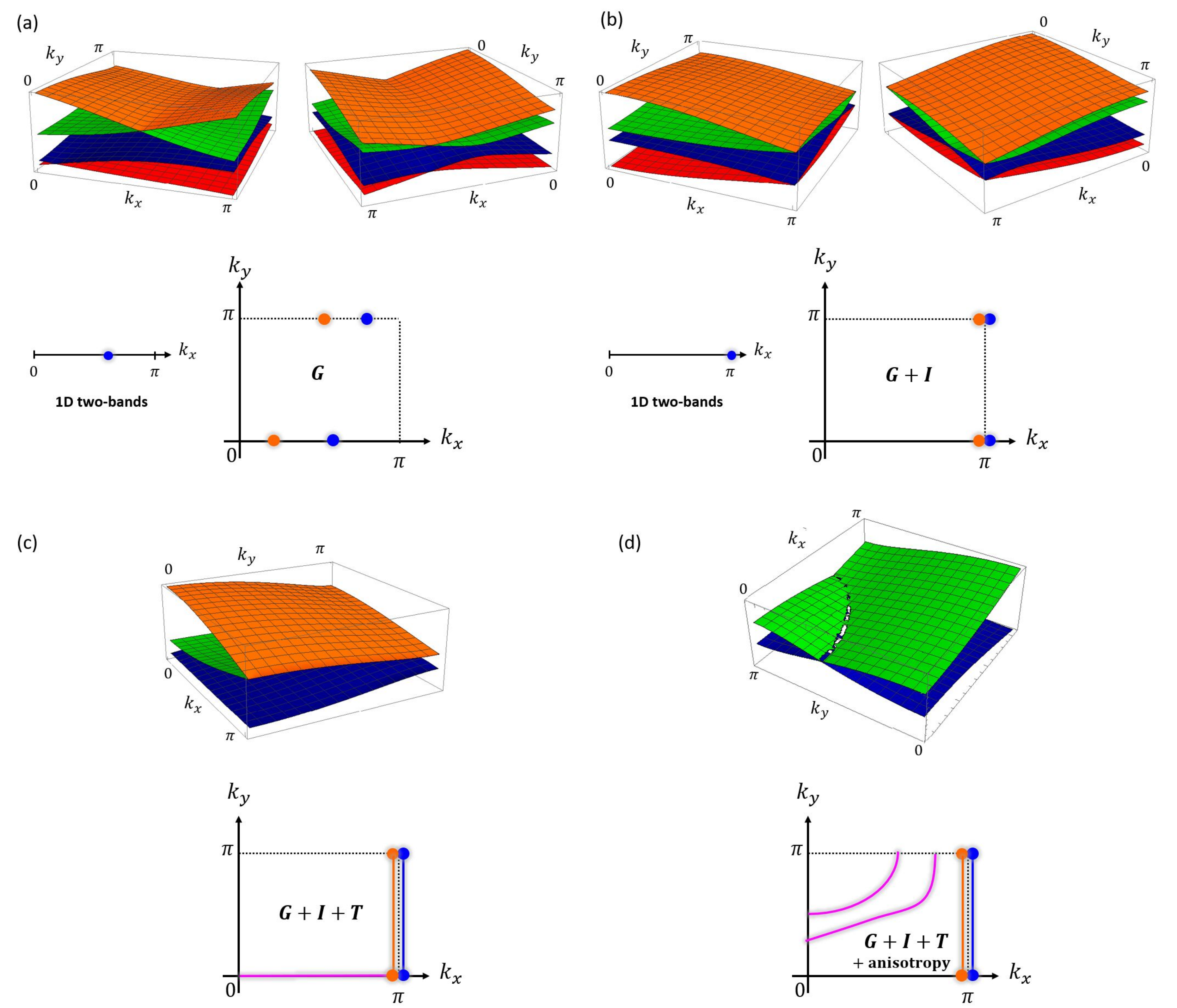}
\caption{(a)-(d) upper panels: Energy bands of the effective model with glide-plane (G) symmetry, with glide-plane (G) + inversion-point (I) symmetries, with glide-plane (G) + inversion-point (I) + time-reversal (T) symmetries, and with glide-plane (G) + inversion-point (I) + time-reversal (T) symmetries + anisotropy, respectively. (a)-(d) lower panels: Projection on the $k_{x}-k_{y}$ plane of the nodal points or nodal lines shown on the corresponding upper panels. The lower panels of (a) and (b) contain also the 1D two-band result for comparison. The lower panel of (d) illustrates two accidental nodal lines inside the Brillouin zone, corresponding to different anisotropic cases.}
\label{Fig4}
\end{figure*}

Instead, we must impose all Eqs.~(\ref{GPconstraints1})-(\ref{GPconstraints4}) on the entries given by Eq.~(\ref{epsilon}). This yields the entries constrained by the glide-plane symmetry given by Eqs.~(\ref{GPconstraint1TB})-(\ref{GPconstraint6TB}) (c.f. Appendix A). The upper panel of Figure~\ref{Fig4}(a) shows two views of the band structure of the glide-plane invariant effective model given by Eqs.~(\ref{GPconstraint1TB})-(\ref{GPconstraint6TB}) on the positive quadrant of the $k_{x}-k_{y}$ plane, for $\mu_{1}=\mu_{3}=0$ and $t_{2,1}=t_{4,3}=t_{1,3}=t_{3,1}=t_{1,4}=t_{4,1}=\text{exp}(i0.3\pi)$ (with $\mu$'s and $t$'s given in arbitrary units). The bands touch pairwise at some value of $k_{x}$, provided $k_{y}=\bar{k}_{y}=0,\pi$. The lower panels of Fig.~\ref{Fig4}(a) depicts the projection of the nodal points on the $k_{x}-k_{y}$ plane, with the 1D two-band model result~\cite{Zhao2016} shown on the left for comparison. Including higher-order hopping in Eq.~(\ref{epsilon}) will move the glide-plane enforced nodal points along $k_{x}$, but will not gap them out since higher-order hopping preserve the glide-plane symmetry. We thus see that exact diagonalization of the effective model confirms the previous prediction derived from the general relations obeyed by the eigenvalues of ${\cal H}(k_{x},k_{y})$ and ${\cal G}(k_{x},k_{y})$, namely, that the glide-plane symmetry enforces two pairs of nodal points, one pair at each line $k_{y}=\bar{k_{y}}=0,\pi$. Next, we show how the remaining symmetries turn the above two pairs of nodal points into two NLs pinned at the $k_{x}=\pi$ edge of the BZ.

\section{Effect of inversion and time-reversal symmetries}

The invariance relation representing the symmetry of the lattice with respect to the inversion-point transformation $I$ is
\begin{equation} \label{IInvariance}
{\cal I}(k_{y}){\cal H}(-k_{x},-k_{y}){\cal I}^{-1}(k_{y})={\cal H}(k_{x},k_{y}),
\end{equation}
where ${\cal I}(k_{y})$ is the $4\times4$ matrix representation of $I$ in the basis constructed by the eigenstates of ${\cal H}(k_{x},k_{y})$. Similar to what happens with ${\cal G}$, the $k_{y}$-dependance of ${\cal I}$ is a consequence of the shift of $I$ from the center of the unit cell along the $m_{y}$-direction (c.f. Fig.~\ref{Fig2}(a)).

Combining the constraints imposed by Eq.~(\ref{IInvariance}) with those from Eq.~(\ref{GInvariance}), we obtain Eqs.~(\ref{GPIPconstraint1TB})-(\ref{GPIPconstraint5TB}) (c.f. Appendix B) which give the off-diagonal entries of the glide-plane and inversion-point invariant effective model. The resulting band structure and projection of nodal points on the $k_{x}-k_{y}$ plane are shown on the upper and lower panels of Fig.~\ref{Fig4}(b) for $\mu_{1}=\mu_{3}=0$, $t_{2,1}=t_{4,3}=1$ ($I$-symmetry implies that these hopping parameters must be real; c.f. Appendix B), and $t_{1,3}=t_{3,1}=t_{1,4}=\text{exp}(i0.3\pi)$. Fig.~\ref{Fig4}(b) conveys that the effect of the inversion-point symmetry is to pin at $k_{x}=\pi$ the four nodal points of the glide-plane invariant effective model. The 1D two-band model result~\cite{Zhao2016} is shown to the left of the lower panel. It can be shown that $G$ and $I$ imply $R$ (c.f. Appendix C). Therefore, it is sufficient to analyze the constraints imposed by $G$ and $I$ on the band structure.

Turning to the time-reversal transformation $T$, the invariance relation is given by
\begin{equation} \label{TInvariance}
{\cal T}{\cal H}^{\ast}(-k_{x},-k_{y}){\cal T}^{-1}={\cal H}(k_{x},k_{y}),
\end{equation}
where ${\cal T}=1\!\!1_{4\times4}$ since ${\cal H}(k_{x},k_{y})$ is spinless.

Imposing Eq. (\ref{TInvariance}) implies that the entries of ${\cal H}(k_{x},k_{y})$ satisfy $\varepsilon^{\ast}_{n,m}(-k_{x},-k_{y})=\varepsilon_{n,m}(k_{x},k_{y})$ which, given Eq.~(\ref{epsilon}), yields real hopping parameters. As can be seen in the upper and lower panels of Fig.~\ref{Fig4}(c), in which $\mu_{1}=\mu_{3}=0$, $t_{2,1}=t_{4,3}=t_{1,3}=t_{3,1}=t_{1,4}=1$, the effect of time-reversal symmetry on the band structure of the glide-plane and inversion-point invariant effective model is two-fold: it connects the symmetry-enforced nodal points of the same pair of bands, thus forming two NLs at $k_{x}=\pi$, and it also induces an additional NL at $k_{y}=0$. While the former NLs are symmetry-enforced, the later is accidental, being gapped, for instance, by an anisotropy of the hopping parameters, as we shall see next.

\section{Effect of anisotropy}

Here we discuss another feature of the band structure which is special to our scheme of engineering a nonsymmorphic symmetry out of vacancies: The appearance of unusually robust accidental NLs in the interior of the BZ.

In a pristine monoatomic lattice in which the distance between neighboring atoms is the same in all directions (such as borophene), the hopping amplitudes are direction-independent. In such an isotropic environment (and disregarding non-structural degrees of freedom such as orbitals and spin), the crossings between energy bands are the ones associated to the symmetries, as we have seen in the previous sections. Defects (including vacancies) break the isotropy of the hopping amplitudes, with the result that now bands can cross also in generic places of the BZ. For the particular band structure shown on Fig.~\ref{Fig4}(c), the anisotropy gaps out the accidental NL at $k_{y}=0$, but creates another accidental NL between the middle bands in the interior of the BZ, as shown on the upper panel of Fig.~\ref{Fig4}(d) where $\mu_{1}=\mu_{3}=0$, $t_{2\,1}=t_{4\,3}=1$, and $t_{1\,3}=t_{3\,1}=t_{1\,4}=100$. Smoothly changing the anisotropy between the hopping amplitudes makes the NL move and change shape through the BZ, as illustrated on the lower panel of Fig.~\ref{Fig4}(d). Eventually, the NL starts to fade and finally disappears when the parameters are taken out of a certain anisotropic regime. Unlike accidental NLs of nonstructural origin, which are easily gapped by perturbations, anisotropy-induced accidental NLs should have an enhanced robustness owing to their structural origin. This feature can be traced to the fact that perturbations do not restore isotropy (sometimes they might actually enhance the anisotropy).

We conclude that a nonsymmorphic 2D material created by vacancy-engineering possesses symmetry-enforced NLs at one edge of the BZ originated from a glide-plane symmetry (combined with inversion-point and time-reversal symmetries), and also accidental NLs with enhanced robustness in the interior of the BZ from the vacancy-induced anisotropy.

\section{Density functional theory results}

The analytical predictions of the previous sections are confirmed by numerical investigations of B$_{10}$ and B$_{16}$ depicted on Figs.~\ref{Fig1}(a)-(b), both as unperturbed sheets and subject to Rashba SOC. The corresponding tight-binding Hamiltonians are presented in Appendix D. The electronic band structures are obtained using density functional theory (DFT) calculations~\cite{Giannozzi2009}, band-unfolding for supercell computations \cite{Medeiros2014}, and tight-binding Hamiltonian computed using the Wannier 90 package~\cite{Mostofi2014}.

\begin{figure*}[htpb]
\centering
\includegraphics[width=16cm]{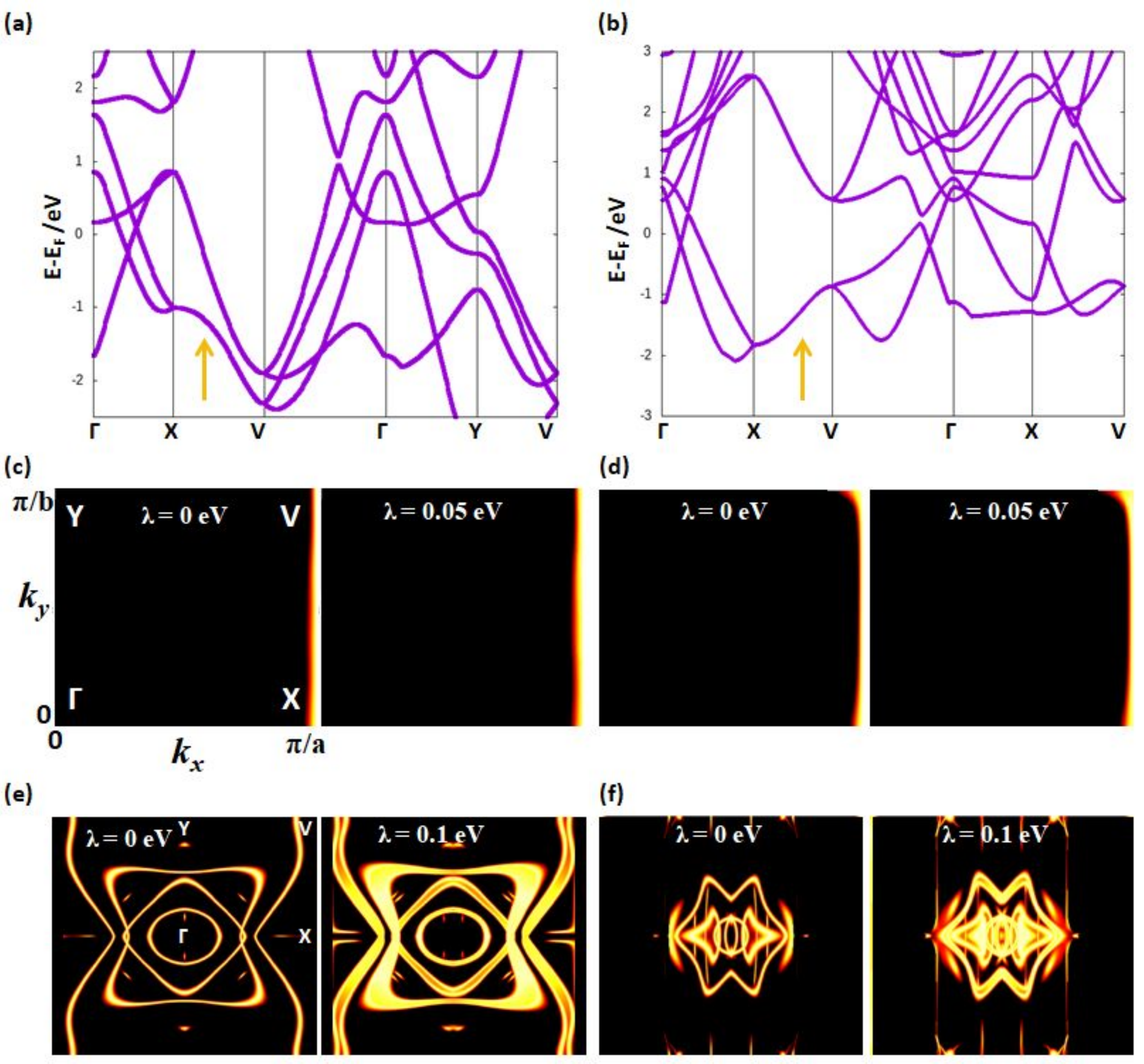}
\caption{(a) [(b)] Band structure of B$_{10}$ [B$_{16}$] without Rashba spin-orbit coupling (SOC) along the $\Gamma$-X-V-$\Gamma$-Y-V path in the Brillouin zone, with the position of points $\Gamma$, X, V, and Y given in the left panel of (c). (c) [(d)] Contour plot of the symmetry-enforced nodal line indicated by the arrow in (a) [(b)] without Rashba SOC in the left panel ($\lambda=0$ eV) and in the presence of Rashba SOC in the right panel ($\lambda=0.05$ eV). (e) [(f)] Contour plot of all accidental nodal lines which exist within an energy window of 2.0 eV about the Fermi energy in the band structure shown in (a) [(b)], without Rashba SOC in the left panel ($\lambda=0$ eV) and in the presence of Rashba SOC in the right panel ($\lambda=0.1$ eV).}
\label{Fig5}
\end{figure*}

Figures~\ref{Fig5}(a)-(b) show the band structure of B$_{10}$ and B$_{16}$, respectively, both without Rashba SOC. The bands are given along the $\Gamma$-X-V-$\Gamma$-Y-V path in the BZ (points $\Gamma$, X, V, and Y are shown on the left panel of Fig.~\ref{Fig5}(c)). Figs.~\ref{Fig5}(a)-(b) indicate that bands stick together pairwise, forming NLs, along the X-V direction (corresponding to $k_{x}=\pi/a$). Figures~\ref{Fig5}(c)-(d), left panels, are the contour plots in the $k_{x}-k_{y}$ plane of the lower NL indicated by the arrow in Figs.~\ref{Fig5}(a)-(b), respectively. In these contour plots, the orange line running along X-V represents the vanishing of the energy difference between the sticking bands. Figures~\ref{Fig5}(e)-(f), left panels, are the contour plots of the NLs which exist in the interior of the BZ of the band structures shown in Figs.~\ref{Fig5}(a)-(b), respectively, within an energy window of 2.0 eV about the Fermi energy. Figs.~\ref{Fig5}(c)-(f), right panels, show the contour plots of the NLs in the corresponding left panes but with Rashba SOC of strength $\lambda=0.05$ eV in Figs.~\ref{Fig5}(c)-(d) and $\lambda=0.1$ eV in Figs.~\ref{Fig5}(e)-(f). 

Rashba SOC preserves the crystalline symmetries, as well as time-reversal symmetry. The NLs shown in Figs.~\ref{Fig5}(c)-(d) correspond to those of the effective model featured on Figs.~\ref{Fig4}(c)-(d) at $k_{x}=\pi$; they are symmetry-enforced, hence their robustness to Rashba SOC. The NLs of Figs.~\ref{Fig5}(e)-(f) are our $\beta$-borophenes' analogues of the NLs shown in Fig.~\ref{Fig4}(d) inside the BZ; despite being accidental, they survive in the regime of very strong Rashba SOC, so they must originate from the vacancy-induced anisotropy. Indeed, as shown in Figs.~\ref{Fig5}(e)-(f), the effect of Rashba SOC of strength as large as $\lambda=0.1$ eV (way beyond the experimental bound) on these accidental NLs is just to lift their spin-degeneracy by shifting the spin-polarized bands in opposite directions in the $k_{x}-k_{y}$ plane.

We note that the phenomena uncovered here are not exclusive to B$_{10}$ and B$_{16}$. We have found other $\beta$-borophenes with concentration of vacancies within the stability range reported in Ref.~\cite{Penev2012}, and with the same symmetries as B$_{10}$ and B$_{16}$~\cite{Liu2021}. These materials thus possess symmetry-enforced NLs and unusually robust accidental NLs in the spectrum.

\section{Outlook}

We have shown how vacancy-engineering turns a monoatomic symmorphic 2D material into a nonsymmorphic one with a glide-plane symmetry. By carrying out a symmetry analysis and applying it to an effective model, we have demonstrated that the synthesized glide plane enforces two pairs of nodal points in the spectrum of the material. When the glide-plane symmetry is combined with inversion-point and time-reversal symmetries, the nodal points of each pair are connected through enforced NLs pinned at one edge of the BZ. We have also uncovered anisotropy-induced accidental NLs which can be moved around in the interior of the BZ by varying the anisotropic parameters, an interesting and potentially very useful byproduct of introducing vacancies. DFT results for vacancy-engineered borophenes confirm the analytical predictions for the enforced NLs, and also convey that these materials have accidental NLs inside the BZ which survive under very strong Rashba SOC. This enhanced robustness is consistent with an anisotropy-origin, and should be contrasted to the usual frailty of common accidental NLs.

In terms of applications, robustness to perturbations is one of the key features of a NLSM. An accidental NL which is robust across the experimentally available ranges of perturbations' strengths is at least as good as a symmetry-enforced one, with the advantage that the former can be moved around in the BZ. Conceivably, the ability to move a NL will affect momentum-dependent scattering processes, or even suppresses them, which can be used to manipulate various susceptibilities of the material. Some guidance in such a design of nodal regions can be inferred from the example of unconventional superconductors: The change in the BZ location of the gap closing modifies the electric, magnetic thermal, and optical responses of the material~\cite{Graf1996,Zutic1997,Prozorov2006}. Finally, our proposal relies only on crystal symmetries, being applicable to general 2D materials, and possibly also to 3D materials. The basic principle of creating a nonsymmorphic symmetry simply by introducing vacancies in a symmorphic crystal thus offers a new path to fundamental investigations and material design of NLSMs and their applications.

\acknowledgments
We thank Anton Burkov and Wei Chen for valuable discussions. This work was supported by CAPES, CNPq, FAPDF (F. L. and F. Q.), and U.S. DOE, Office of Science BES, Award No. DE-SC0004890 (I. \v{Z}.).

\appendix

\section{Appendix A: Glide-plane symmetry}

\renewcommand{\theequation}{A\thesection.\arabic{equation}}

\setcounter{equation}{0}

Let $|k_{x},k_{y}\rangle_{\alpha}$ ($\alpha=1,2,3,4$) be the eigenstate of the Bloch Hamiltonian ${\cal H}(k_{x},k_{y})$ with eigenenergy $E_{\alpha}(k_{x},k_{y})$, i.e.
\begin{equation} \label{EigenstatesAndEigenvalues}
{\cal H}(k_{x},k_{y})|k_{x},k_{y}\rangle_{\alpha}=E_{\alpha}(k_{x},k_{y})|k_{x},k_{y}\rangle_{\alpha}.
\end{equation}

Combining Eqs.~(\ref{GInvariance}) and (\ref{EigenstatesAndEigenvalues}) we arrive at
\begin{eqnarray} \label{Relation}
\nonumber &{\cal H}&(k_{x},k_{y})\,\,{\cal G}(k_{x},k_{y})|k_{x},-k_{y}\rangle_{\alpha'}=\\
&=&E_{\alpha'}(k_{x},-k_{y})\,\,{\cal G}(k_{x},k_{y})|k_{x},-k_{y}\rangle_{\alpha'},
\end{eqnarray}
which means that ${\cal G}(k_{x},k_{y})|k_{x},-k_{y}\rangle_{\alpha'}$ is an eigenstate of ${\cal H}(k_{x},k_{y})$ with eigenenergy $E_{\alpha'}(k_{x},-k_{y})$. It follows that ${\cal G}(k_{x},k_{y})|k_{x},-k_{y}\rangle_{\alpha'}$ equals an eigenstate $|k_{x},k_{y}\rangle_{\alpha}$ up to a phase, i.e.,
\begin{equation} \label{GlideTransformationOfEigenstates}
{\cal G}(k_{x},k_{y})|k_{x},-k_{y}\rangle_{\alpha'}=e^{i\theta_{\alpha',\alpha}(k_{x},k_{y})}|k_{x},k_{y}\rangle_{\alpha}.
\end{equation}
It also follows that $E_{\alpha'}(k_{x},-k_{y})=E_{\alpha}(k_{x},k_{y})$.

On the lines $k_{y}=\bar{k}_{y}=0,\pm\pi$, ${\cal H}(k_{x},-\bar{k}_{y})={\cal H}(k_{x},\bar{k}_{y})$ (where ${\cal H}(k_{x},\mp\pi)={\cal H}(k_{x},\pm\pi)$ follows from the $2\pi$-periodicity of the BZ). On these lines, Eq.~(\ref{GInvariance}) yields $[{\cal G}(k_{x},\bar{k}_{y}),{\cal H}(k_{x},\bar{k}_{y})]=0$, and hence ${\cal G}(k_{x},\bar{k}_{y})$ and ${\cal H}(k_{x},\bar{k}_{y})$ share a set of eigenstates. Indeed, substituting $k_{y}$ by $\bar{k}_{y}$ in Eq.~(\ref{Relation}) (and using that ${\cal H}(k_{x},\bar{k}_{y})={\cal H}(k_{x},-\bar{k}_{y})$), we conclude that ${\cal G}(k_{x},\bar{k}_{y})|k_{x},\bar{k}_{y}\rangle_{\alpha'}$ is an eigenstate of ${\cal H}(k_{x},\bar{k}_{y})$ with eigenenergy $E_{\alpha'}(k_{x},\bar{k}_{y})$. Since the eigenenergies are, in general, non-degenerate (i.e., $E_{\alpha}(k_{x},k_{y}) \neq E_{\alpha'}(k_{x},k_{y})$ for $\alpha \neq \alpha'$), it follows that ${\cal G}(k_{x},\bar{k}_{y})|k_{x},\bar{k}_{y}\rangle_{\alpha'}$ equals $|k_{x},\bar{k}_{y}\rangle_{\alpha'}$ up to a phase, i.e.,
\begin{equation} \label{EigenstatesOfG}
{\cal G}(k_{x},\bar{k}_{y})|k_{x},\bar{k}_{y}\rangle_{\alpha'}=e^{i\theta_{\alpha'}(k_{x})}|k_{x},\bar{k}_{y}\rangle_{\alpha'}.
\end{equation}

The matrix ${\cal G}(k_{x},k_{y})$ can be obtained by extracting how the glide plane $G$ transforms the second-quantized operator $c_{m_{x},m_{y}}^{j}$ acting on the $j$-th site of the unit cell located at $(m_{x},m_{y})$. Using Fig.~\ref{Fig2}(a), it is easy to see that $G$ acts on the operator of the unit cell at $(-1,-1)$ as
\begin{eqnarray}
\nonumber G\,c_{-1,-1}^{1}=c_{-1,0}^{2} \qquad\qquad G\,c_{-1,-1}^{2}=c_{0,0}^{1}\\
\nonumber G\,c_{-1,-1}^{3}=c_{-1,1}^{4} \qquad\qquad G\,c_{-1,-1}^{4}=c_{0,1}^{3}
\end{eqnarray}

For an unit cell at $(-m_{x},-m_{y})$, $G$ act as
\begin{eqnarray}
\nonumber G\,c_{-m_{x},-m_{y}}^{1}=\,&c&_{-m_{x},m_{y}-1}^{2} \qquad G\,c_{-m_{x},-m_{y}}^{2}=c_{-m_{x}+1,m_{y}-1}^{1}\\
\nonumber G\,c_{-m_{x},-m_{y}}^{3}=\,&c&_{-m_{x},m_{y}}^{4} \qquad\quad G\,c_{-m_{x},-m_{y}}^{4}=c_{-m_{x}+1,m_{y}}^{3}.
\end{eqnarray}

Applying the Fourier transform of $c_{m_{x},m_{y}}^{j}$, i.e.
\begin{equation}
\nonumber c_{m_{x},m_{y}}^{j} = \sum_{k_{x},k_{y}}c_{k_{x},k_{y}}^{j}e^{i(k_{x}m_{x}+k_{y}m_{y})},
\end{equation}
we obtain these relations in momentum space as
\begin{eqnarray}
\nonumber G\,c_{k_{x},k_{y}}^{1}=&e&^{ik_{y}}c_{k_{x},-k_{y}}^{2} \qquad\quad G\,c_{k_{x},k_{y}}^{2}=e^{i(k_{x}+k_{y})}c_{k_{x},-k_{y}}^{1}\\
\nonumber G\,c_{k_{x},k_{y}}^{3}=&c&_{k_{x},-k_{y}}^{4} \qquad\qquad\quad G\,c_{k_{x},k_{y}}^{4}=e^{ik_{x}}c_{k_{x},-k_{y}}^{3}.
\end{eqnarray}

The above transformations can be carried out by applying the operator ${\cal G}(k_{x},k_{y})\Updownarrow_{k_{y}}$ to the spinor $c_{k_{x},k_{y}}=[c_{k_{x},k_{y}}^{1}\,\,\,\,c_{k_{x},k_{y}}^{2}\,\,\,\,c_{k_{x},k_{y}}^{3}\,\,\,\,c_{k_{x},k_{y}}^{4}]^{T}$, where $\Updownarrow_{k_{y}}$ flips $k_{y}$ and ${\cal G}(k_{x},k_{y})$ is the $4\times4$ matrix
\begin{equation}\label{G}
{\cal G}(k_{x},k_{y}) =  \begin{bmatrix}
    e^{ik_{y}}g(k_{x}) & 0\\
    0 & g(k_{x})\\
\end{bmatrix}, \qquad g(k_{x}) =  \begin{bmatrix}
    0 & 1\\
    e^{ik_{x}} & 0\\
\end{bmatrix}.
\end{equation}

The eigenvalues of ${\cal G}(k_{x},\bar{k}_{y})$ are found by solving the characteristic equation
\begin{widetext}
\begin{equation}
\nonumber \text{det}[{\cal G}(k_{x},\bar{k}_{y})-\xi1\!\!1]=\text{det}\begin{bmatrix}
    \pm g(k_{x})-\xi1\!\!1 & 0\\
    0 & g(k_{x})-\xi1\!\!1\\
\end{bmatrix}=\text{det}[\pm g(k_{x})-\xi1\!\!1]\text{det}[g(k_{x})-\xi1\!\!1]=(\xi^2-e^{ik_{x}})^2=0.
\end{equation}
\end{widetext}

The two-fold degenerate eigenvalues of ${\cal G}(k_{x},\bar{k}_{y})$ are thus
\begin{equation}\label{EigenvaluesOfG}
\xi_{1,3}=e^{ik_{x}/2},\,\,\xi_{2,4}=-e^{ik_{x}/2}.
\end{equation}

By enforcing Eq. (\ref{GInvariance}), with ${\cal G}(k_{x},k_{y})$ given by Eq.(\ref{G}), the entries $\varepsilon_{n,m}(k_{x},k_{y})$ of ${\cal H}(k_{x},k_{y})$ get constrained by the relations:
\begin{equation}\label{GPconstraints1}
\varepsilon_{n+1,n+1}\,=\,\varepsilon_{n,n},\, n=1,3;
\end{equation}
\begin{equation}\label{GPconstraints2}
\varepsilon_{n,n+1}^{\ast}(k_{x},k_{y})\,=\,e^{ik_{x}}\,\varepsilon_{n,n+1}(k_{x},-k_{y}),\, n=1,3;
\end{equation}
\begin{equation}\label{GPconstraints3}
\varepsilon_{2,3}(k_{x},k_{y})\,=\,e^{i(k_{x}+k_{y})}\,\varepsilon_{1,4}(k_{x},-k_{y});
\end{equation}
\begin{equation}\label{GPconstraints4}
\varepsilon_{2,4}(k_{x},k_{y})\,=\,e^{ik_{y}}\,\varepsilon_{1,3}(k_{x},-k_{y}).
\end{equation}

For the tight-binding model whose entries are given by Eq.~(\ref{epsilon}) (with the specified conditions for the parameters given below Eq.~(\ref{epsilon})), Eqs.~(\ref{GPconstraints1})-(\ref{GPconstraints4}) yield
\begin{equation}\label{GPconstraint1TB}
\varepsilon_{n+1,n+1}\,=\,\varepsilon_{n,n}\,=\,2\mu_{n},\, n=1,3;
\end{equation}
\begin{equation}\label{GPconstraint2TB}
\varepsilon_{n,n+1}(k_{x},k_{y})\,=\,2i\text{Im}(t_{n+1,n})+t^{\ast}_{n+1,n}(1+e^{-ik_{x}}),\, n=1,3;
\end{equation}
\begin{equation}\label{GPconstraint3TB}
\varepsilon_{1,3}(k_{x},k_{y})\,=\,t_{3,1}^{\ast}+t_{1,3}(1+e^{ik_{y}});
\end{equation}
\begin{equation}\label{GPconstraint4TB}
\varepsilon_{1,4}(k_{x},k_{y})\,=\,t_{1,4}+t^{\ast}_{4,1}(1+e^{-ik_{x}});
\end{equation}
\begin{equation}\label{GPconstraint5TB}
\varepsilon_{2,3}(k_{x},k_{y})\,=\,e^{i(k_{x}+k_{y})}\,[t_{1,4}+t^{\ast}_{4,1}(1+e^{-ik_{x}})];
\end{equation}
\begin{equation}\label{GPconstraint6TB}
\varepsilon_{2,4}(k_{x},k_{y})\,=\,e^{ik_{y}}\,[t_{3,1}^{\ast}+t_{1,3}(1+e^{-ik_{y}})];
\end{equation}
and $\varepsilon_{m,n}(k_{x},k_{y})=\varepsilon^{\ast}_{n,m}(k_{x},k_{y})$.

\section{Appendix B: Inversion-point symmetry}

\renewcommand{\theequation}{B\thesection.\arabic{equation}}

\setcounter{equation}{0}

A similar procedure to the one outlined in Appendix A for $G$ yields the operator ${\cal I}(k_{y})\Updownarrow_{k_{x},k_{y}}$ describing the inversion-point transformation $I$, with
\begin{equation}\label{I}
{\cal I}(k_{y}) =  \begin{bmatrix}
    e^{ik_{y}}\sigma_{x} & 0\\
    0 & \sigma_{x}\\
\end{bmatrix}, \qquad \sigma_{x} =  \begin{bmatrix}
    0 & 1\\
    1 & 0\\
\end{bmatrix}.
\end{equation}

The constraints imposed by Eq. (\ref{IInvariance}), with ${\cal I}(k_{y})$ given by Eq.(\ref{I}), on the entries $\varepsilon_{n,m}(k_{x},k_{y})$ of ${\cal H}(k_{x},k_{y})$ are
\begin{equation}\label{IPconstraints1}
\varepsilon_{n,n+1}^{\ast}(k_{x},k_{y})\,=\,\varepsilon_{n,n+1}(-k_{x},-k_{y}),\, n=1,3;
\end{equation}
\begin{equation}\label{IPconstraints2}
\varepsilon_{2,3}(k_{x},k_{y})\,=\,e^{ik_{y}}\,\varepsilon_{1,4}(-k_{x},-k_{y});
\end{equation}
\begin{equation}\label{IPconstraints3}
\varepsilon_{2,4}(k_{x},k_{y})\,=\,e^{ik_{y}}\,\varepsilon_{1,3}(-k_{x},-k_{y}).
\end{equation}

Applying Eqs.~(\ref{IPconstraints1})-(\ref{IPconstraints3}) to the tight-binding model with entries given by Eq.~(\ref{epsilon}) (with the specified conditions for the parameters given below Eq.~(\ref{epsilon})) leads to
\begin{equation}\label{IPconstraint1TB}
\varepsilon_{n,n+1}(k_{x},k_{y})\,=\,t_{n,n+1}+t_{n+1,n}(1+e^{-ik_{x}}),\, n=1,3;
\end{equation}
\begin{equation}\label{IPconstraint2TB}
\varepsilon_{1,3}(k_{x},k_{y})\,=\,t_{3,1}^{\ast}+t_{1,3}(1+e^{ik_{y}});
\end{equation}
\begin{equation}\label{IPconstraint3TB}
\varepsilon_{1,4}(k_{x},k_{y})\,=\,t_{1,4}+t^{\ast}_{4,1}(1+e^{-ik_{x}});
\end{equation}
\begin{equation}\label{IPconstraint4TB}
\varepsilon_{2,3}(k_{x},k_{y})\,=\,e^{ik_{y}}\,[t_{1,4}+t^{\ast}_{4,1}(1+e^{ik_{x}})];
\end{equation}
\begin{equation}\label{IPconstraint5TB}
\varepsilon_{2,4}(k_{x},k_{y})\,=\,e^{ik_{y}}\,[t_{3,1}^{\ast}+t_{1,3}(1+e^{-ik_{y}})],
\end{equation}
with $t_{n,n+1}$ and $t_{n+1,n}\in\Re$, $n=1,3$, and $\varepsilon_{m,n}(k_{x},k_{y})=\varepsilon^{\ast}_{n,m}(k_{x},k_{y})$.

Pairwise combining Eqs.~(\ref{GPconstraint2TB})-(\ref{GPconstraint6TB}) and Eqs.~(\ref{IPconstraint1TB})-(\ref{IPconstraint5TB}) yields the off-diagonal entries of the glide-plane and inversion-point invariant tight-binding model:
\begin{equation}\label{GPIPconstraint1TB}
\varepsilon_{n,n+1}(k_{x},k_{y})\,=\,t_{n+1,n}(1+e^{-ik_{x}});
\end{equation}
\begin{equation}\label{GPIPconstraint2TB}
\varepsilon_{1,3}(k_{x},k_{y})\,=\,t_{3,1}^{\ast}+t_{1,3}(1+e^{ik_{y}});
\end{equation}
\begin{equation}\label{GPIPconstraint3TB}
\varepsilon_{1,4}(k_{x},k_{y})\,=\,t^{\ast}_{4,1}(1+e^{-ik_{x}});
\end{equation}
\begin{equation}\label{GPIPconstraint4TB}
\varepsilon_{2,3}(k_{x},k_{y})\,=\,e^{ik_{y}}\,[t^{\ast}_{4,1}(1+e^{ik_{x}})];
\end{equation}
\begin{equation}\label{GPIPconstraint5TB}
\varepsilon_{2,4}(k_{x},k_{y})\,=\,e^{ik_{y}}\,[t_{3,1}^{\ast}+t_{1,3}(1+e^{-ik_{y}})],
\end{equation}
with $t_{n+1,n}\in\Re$, $n=1,3$. The diagonal entries are given by Eq.~(\ref{GPconstraint1TB}).

\section{Appendix C: Reflection-plane symmetry}

\renewcommand{\theequation}{C\thesection.\arabic{equation}}

\setcounter{equation}{0}

A similar procedure to the one outlined in Appendix A for $G$ yields the operator ${\cal R}(k_{x})\Updownarrow_{k_{x}}$ describing the reflection-plane transformation $R$, with
\begin{equation}\label{R}
{\cal R}(k_{x})={\cal G}(k_{x},k_{y}){\cal I}(-k_{y}),
\end{equation}
and ${\cal G}(k_{x},k_{y})$ and ${\cal I}(k_{y})$ given by Eqs.~(\ref{G}) and (\ref{I}), respectively.

Rewriting Eq.~(\ref{IInvariance}) as ${\cal I}(-k_{y}){\cal H}(-k_{x},k_{y}){\cal I}^{-1}(-k_{y})={\cal H}(k_{x},-k_{y})$ and inserting that into Eq.~(\ref{GInvariance}), we get ${\cal G}(k_{x},k_{y}){\cal I}(-k_{y}){\cal H}(-k_{x},k_{y}){\cal I}^{-1}(-k_{y}){\cal G}^{-1}(k_{x},k_{y})={\cal H}(k_{x},k_{y})$. Now, using Eq.~({\ref{R}}), the latter expression yields ${\cal R}(k_{x}){\cal H}(-k_{x},k_{y}){\cal R}^{-1}(k_{x})={\cal H}(k_{x},k_{y})$, which is the invariance relation of ${\cal H}(k_{x},k_{y})$ with respect to ${\cal R}(k_{x})$. That is, if the glide plane $G$ and the inversion point $I$ are symmetries, then the reflection plane $R$ is also a symmetry.

\section{Appendix D: Tight-binding model used in the\\
density functional theory calculations}

\renewcommand{\theequation}{E\thesection.\arabic{equation}}

\setcounter{equation}{0}

Let ${\bf R}_{n,m}$ be the position of the unit cell at the point $(m, n)$ of the Bravais lattice, and ${\bf r}_{\tau}$ the position of the $\tau$-atom in the unit cell. The matrix entries of the tight-binding Hamiltonian $H_{TB}$ are given by
\begin{eqnarray}\label{HTBentries}
\nonumber H_{\tau\alpha,\tau'\alpha'}({\bf k})&=&\varepsilon_{\tau\alpha}\delta_{\alpha,\alpha'}\delta_{\tau,\tau'}+\\
\nonumber &+&\sum_{n}\sum_{m}e^{i{\bf k}\cdot({{\bf R}_{n,m}+{\bf r}_{\tau}-{\bf r}_{\tau'})}}t_{\alpha,\alpha'}({\bf R}_n+{\bf r}_{\tau}-{\bf r}_{\tau'}),
\end{eqnarray}
where $t_{\alpha,\alpha'}({\bf R}_{n}+{\bf r}_{\tau}- {\bf r}_{\tau'})$ is the hopping integral between the $\alpha$-orbital of the $\tau$-atom at ${\bf r}_{\tau}$ and the $\alpha'$-orbital of the $\tau{'}$-atom at ${\bf r}_{\tau'}$, and $\varepsilon_{\tau,\alpha}$ is the atomic energy of the $\alpha$-orbital of the $\tau$-atom.

Rashba SOC is incorporated by adding $H_{SOC}$ to $H_{TB}$ with
\begin{equation}\label{HSOC}
\nonumber H_{SOC}=i\sum_{n}\lambda_{n}\sum_{\langle{i,j}\rangle}c_{i}^{+}[\sigma\times{\bf d}_{i,j}^{n}]c_{j},
\end{equation}
where $c_{i}^{+}$ ($c_{i}$) is the creation (annihilation) operator of an electron at site $i=(m,n)$ of the Bravais lattice, ${\bf d}_{i,j}^{n}={\bf r}_{i}-{\bf r}_{j}$ is the distance connecting the $n$-th nearest-neighbor sites $i$ and $j$, and $\lambda_{n}$ is the strength of the SOC between the $n$-th nearest-neighbor sites.

\end{document}